\documentclass[prb,twocolumn]{revtex4}

\usepackage{graphicx}
\usepackage{dcolumn}
\usepackage{bm}

\begin{document}

\preprint{APS/123-QED}

\title{Quantum kinetic description of Coulomb effects\\
in one-dimensional nano-transistors}

\author{K. M. Indlekofer}
\email{m.indlekofer@fz-juelich.de}
\author{J. Knoch}
\affiliation{
Institute for Thin Films and Interfaces (ISG-1) and
Center of Nanoelectronic Systems for Information Technology (CNI),
Research Centre J\"ulich GmbH, D-52425 J\"ulich, Germany
}

\author{J. Appenzeller}
\affiliation{
IBM T. J. Watson Research Center, P.O. Box 218,
Yorktown Heights, New York 10598
}

\date{\today}

\begin{abstract}
In this article, we combine the modified electrostatics of a
one-dimensional transistor structure with a quantum kinetic
formulation of Coulomb interaction and nonequilibrium transport.
A multi-configurational self-consistent Green's function approach is
presented, accounting for fluctuating electron numbers.
On this basis we provide a theory for the simulation of electronic
transport and quantum charging effects in nano-transistors,
such as gated carbon nanotube and whisker devices
and one-dimensional CMOS transistors.
Single-electron charging effects arise naturally as a consequence of the
Coulomb repulsion within the channel.
\end{abstract}

\pacs{XXX}
\maketitle

\section{\label{sec:intro}Introduction}

As scaling of field-effect-transistor (FET) devices reaches the
deca-nanometer regime, multi-gate architectures and ultrathin
active channel regions are mandatory in order to preserve
electrostatic integrity. It has been shown that a coaxially gated
nanowire represents the ideal device structure for ultimately
scaled FETs.\cite{aut97,won02}
A variety of 1D nanostructures - such as carbon nanotubes,
silicon nanowires or compound semiconductor nano-whiskers - have
been demonstrated and intensive research has been devoted to the
realization of field-effect-transistor action in these
nanostructures.\cite{jav04,lin05,alp03,the03}
Due to the small lateral
extent in the nanometer range, electronic transport through such
nanowires is one-dimensional with only a few or even a single
transverse mode participating in the current.
As a result, increasingly less
electrons are involved in the switching of a nanowire
transistor. In fact, even in devices with rather long channel
lengths of 100nm, only on the order of 1-10 electrons constitute the
charge in the channel for on-state voltage conditions.
Hence, single-electron charging
effects are increasingly important and have to be taken into
account.\cite{yon01,suz02,aml03}

Two different approaches are commonly used to describe electronic transport
in nano-transistors:
A quantum kinetic approach based on real-time Green's functions
provides an excellent description of non-equilibrium states.
\cite{lak97,yong02,yong01}
Here, the Coulomb interaction is described in terms of a
selfconsistent Hartree potential,
optionally combined with a spin-density-functional
exchange-correlation term in local density approximation (LDA-SDFT).
However, this framework does not account for single-electron charging effects
without forcing integer electron numbers.
Alternatively, the second approach considers a quasi-isolated nanosystem
with a many-body formulation of Coulomb interaction,
including electronic transport on a basis of rate-equations.
\cite{ben91,ave91,jov93,wei95,tan96,pfa95,ind00}
While predicting single-electron charging effects correctly,
the latter neglects dissipation and renormalization effects
due to the source and drain contacts.

Here, we present a novel approach that allows to combine
a quantum kinetic description of non-equilibrium electron transport
with non-local many-body Coulomb effects in one-dimensional FET nanodevices.
Within our approach, single-electron charging effects arise naturally
as a consequence of the Coulomb interaction.
Our formalism contains two central ingredients:
In order to cope with particle-number fluctuations under nonequilibrium
conditions, we introduce a multi-configurational self-consistent Green's
function algorithm.
Secondly, we consider a one-dimensional Coulomb Green's
function for the transistor channel that allows to properly incorporate
many-body interaction effects into a quantum kinetic approach
with electrostatic boundary conditions for a realistic FET.
As an example, we calculate the transfer characteristics of
a nanowire transistor with Schottky-barriers (SB) at the
contact-channel interfaces.

\section{\label{sec:coul}Coulomb Green's function}

\begin{figure}[ht]
\vspace{-0.0cm}
\begin{center}
\includegraphics[width=7cm]{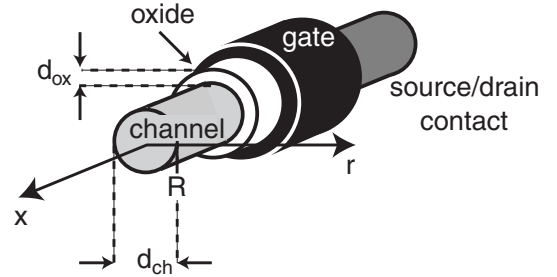}
\end{center}
\vspace{-0.5cm}
\caption{\footnotesize
\label{fig:coax}
Schematic view of a 1D FET geometry.
($d_{ox}$ and $d_{ch}$ denote the gate insulator thickness
and the channel diameter of the nano-transistor, respectively.)
}
\end{figure}

Consider a coaxially gated field-effect-transistor as
illustrated in Fig.~\ref{fig:coax}. A cylindrical semiconducting
channel material is surrounded by a thin dielectric and a metallic
gate electrode.
The electrostatic potential $V$ inside such a one-dimensional (1D)
transistor channel
obeys a modified Poisson equation \cite{aut97,pik97}
\begin{equation}
\label{eq:modpoisson}
\frac{\partial^2}{\partial x^2}V(x)-\frac{1}{\lambda^2}V(x)
=-\frac{1}{\epsilon_0\epsilon_{ch} S}\rho(x)-\frac{1}{\lambda^2}V_G,
\end{equation}
where $\rho$ is the 1D charge density.
$V_G$ denotes the gate potential and $S=\pi d^2_{ch}/4$ is the effective
cross-sectional area.
The characteristic length $\lambda$ is related to the spatial
separation of the gate electrode from the channel (which should be
smaller than the total length $L$ of the channel).\cite{aut97,pik97}
Note that Eq.~(\ref{eq:modpoisson})
is an appropriate description for coaxial as well as planar
transistor geometries, differing only in the characteristic length $\lambda$.
In the following, we assume perfect metallic source and drain
contacts at the boundaries,
yielding fixed-potential boundary conditions
due to given chemical potentials within these contacts.

A key ingredient in our formalism is the usage of a Coulomb Green's function
for the description of charge interaction within the channel.
This allows us to formulate classical electrostatics
(with boundary conditions) and many-body interaction between electrons
on equal footing.
The corresponding Coulomb Green's function
of the gated channel
(with $0\leq x,x'\leq L$ and vanishing potential at the boundaries $0,L$)
can readily be obtained as
\begin{eqnarray}
\label{eq:vgreen}
v(x,x')
&=&
\frac{\lambda}{2}\left(
e^{-\frac{|x-x'|}{\lambda}}-e^{-\frac{x+x'}{\lambda}}
\right)
\\\nonumber
&&+
\frac{\lambda}{2}~e^{-\frac{L}{\lambda}}\left(
\frac{\cosh\left(\frac{x-x'}{\lambda}\right)}
     {\sinh\left(\frac{L}{\lambda}\right)}
-
\frac{\cosh\left(\frac{x+x'}{\lambda}\right)}
     {\sinh\left(\frac{L}{\lambda}\right)}
\right).
\end{eqnarray}
(In contrast, if we considered open boundary conditions, we would obtain
$v(x,x')=(\lambda/2)\exp\left(-|x-x'|/\lambda\right)$ instead.)
For a given charge density $\rho$ inside the channel,
the potential thus reads
\begin{equation}
\label{eq:vpot}
V(x)=\frac{1}{\epsilon_0\epsilon_{ch} S}\!\int\!\!dx'~v(x,x')~\rho(x')
+V_{ext}(x),
\end{equation}
with the external potential contribution
\begin{eqnarray}
\label{eq:vext}
V_{ext}(x)&=&
\frac{\sinh\left(\frac{L-x}{\lambda}\right)}
     {\sinh\left(\frac{L}{\lambda}\right)}V_S
+
\frac{\sinh\left(\frac{x}{\lambda}\right)}
     {\sinh\left(\frac{L}{\lambda}\right)}V_D
\\\nonumber
&&+
\frac{1}{\lambda^2}\!\int\!\!dx'~v(x,x')~V_G,
\end{eqnarray}
where $V_{S}$ and $V_{D}$ denote the contact potentials.

\section{\label{sec:hamop}System Hamiltonian}

In this article, we make use of a tight-binding description of the system,
represented by a localized 1D single-particle basis $\{\phi_j(x,\sigma)\}$
(where the single-particle index $j$ represents
a combined orbital, site, and spin multi-index.)
The total system Hamiltonian $H=H_0+H_{ee}+H_S+H_D$
can be split into four parts.
$H_0$ contains all single-particle terms of the transistor channel:
\begin{eqnarray}
\label{eq:hamop0}
H_0&=&\sum_{i,j}h_{jk}~ c^{\dagger}_jc_k,
\\\nonumber
h_{jk}&=&
-e\sum_{\sigma}\!\!\int\!\!dx
~\phi^*_j(x,\sigma)~\phi_k(x,\sigma)
\Big[V_{dop}(x)+V_{ext}(x)\Big]
\\\nonumber
&&
+\delta_{jk}d_j+t_{jk},
\end{eqnarray}
with the electron annihilation operator $c_j$ for state $j$.
The composition of the channel
(material-specific properties, layer sequence, etc.) is described by $d_j$
and off-diagonal hopping matrix elements $t_{jk}$.\cite{vog83,sto94}
$V_{dop}$ denotes the potential resulting from fixed charges $\rho_{dop}$
(due to ionized doping atoms), whereas $V_{ext}$
stems from external charges due to the applied drain-source voltage
and the gate influence (see Eq.~(\ref{eq:vext})).

Furthermore, $H_S$ and $H_D$ are the Hamiltonians for the source and drain
contacts, respectively.
Latter also contain the corresponding
hopping terms to the outer ends of the channel region,
providing electron injection and absorption.
Each contact is assumed to be in a state of local equilibrium
with an individual chemical potential according to the applied voltage.
(See also Eqs.~(\ref{eq:dyson}),~(\ref{eq:couplself}) below.)

Most importantly, $H_{ee}$ describes the many-body Coulomb interaction
between electrons within the channel region:\cite{ind00}
\begin{eqnarray}
\label{eq:hamopc}
H_{ee}=\frac{1}{2}\sum_{i,j,k,l}V_{mjkl}~c^{\dagger}_mc^{\dagger}_jc_kc_l,
\end{eqnarray}
with Coulomb matrix elements
\begin{eqnarray}
\label{eq:vmatrix}
V_{mjkl}&=&\frac{e^2}{\epsilon_0\epsilon_{ch} S}
\sum_{\sigma,\sigma'}\int\!\!dx\!\!\int\!\!dx'~v(x,x')
\\\nonumber
&&\times
~\phi^*_m(x,\sigma)~\phi^*_j(x'\!,\sigma')
~\phi_k(x'\!,\sigma')~\phi_l(x,\sigma),
\end{eqnarray}
which employ the Coulomb Green's function Eq.~(\ref{eq:vgreen}).

\section{\label{sec:qkin}Quantum kinetics}

A quantum kinetic description of the system (under nonequilibrium
conditions in particular) is obtained via the usage of a
real-time Green's functions formalism.\cite{schaefer,haugjauho,datta}
The retarded and lesser (two-point) Green's functions in the time domain
are given by
\begin{eqnarray}
G^{r}_{jk}(t)&=&-i\Theta(t)\langle
\big\{c_j(t),c^{\dagger}_k(0)\big\}\rangle,\\\nonumber
G^{<}_{jk}(t)&=&i\langle c^{\dagger}_k(0)c_j(t)\rangle,
\end{eqnarray}
for steady-state conditions.
In the following, we will consider the Fourier transformed functions,
defined via $G(E)=(1/\hbar)\!\int\!\!dt\exp(iEt/\hbar)G(t)$.

For temperatures $T$ well above the Kondo temperature of the system,
the Coulomb interaction can be treated independently
of the contact coupling, albeit self-consistently.
In matrix notation, the Dyson equation for the channel thus can be written as
\cite{datta,hen94,lak97}
\begin{eqnarray}
\label{eq:dyson}
G^{r}&=&G^{r0}+G^{r0}\left[
\Sigma^{r}_{ee}+\Sigma^{r}_S+\Sigma^{r}_D
\right]G^{r},\\\nonumber
G^{<}&=&i~f_0~A\\\nonumber
&&+i~(f_S-f_0)~G^{r}\Gamma_S G^{a}\\\nonumber
&&+i~(f_D-f_0)~G^{r}\Gamma_D G^{a},
\end{eqnarray}
with $\Gamma_S\equiv i(\Sigma^{r}_S-\Sigma^{a}_S)$,
$\Gamma_D\equiv i(\Sigma^{r}_D-\Sigma^{a}_D)$
and $A\equiv i(G^{r}-G^{a})$.
$f_S$ and $f_D$ are the local source and drain Fermi distribution functions,
respectively, assuming local equilibrium within these reservoirs.
On the other hand, $f_0$ denotes the equilibrium distribution function of
the isolated channel system (typically, $f_0=f_D$).
Furthermore, $G^{r0}\equiv (E-h+i\epsilon)^{-1}$
(with $\epsilon\to 0+$) and $G^{a}={G^{r}}^\dagger$.

Once $G^{<}$ has been obtained selfconsistently from Eq.~(\ref{eq:dyson}),
observables like the
electron density $\rho_e$ and the current $I_e$
(through an arbitrary layer at $x_0$) can be calculated via
\begin{eqnarray}
\label{eq:erho}
\rho_e(x,\sigma)&=&-e\sum_{jk}
~\phi^*_j(x,\sigma)~\phi_k(x,\sigma)~\hat{\rho}_{jk},
\\\nonumber
I_e&=&-\frac{e}{\hbar}
\!\!\!\sum_{j,k \atop {x_j\leq x_0, \atop x_k>x_0}}\!\!\!
2~Im\left(t_{jk}~\hat{\rho}_{jk}\right),
\end{eqnarray}
with the single-particle density-matrix
\begin{equation}
\label{eq:densma}
\hat{\rho}_{jk}=\frac{1}{2\pi}\!\!\int\!\!dE~\frac{1}{i}G^{<}_{kj}(E).
\end{equation}

The effective contact selfenergies due to the coupling of the channel
to the source and drain regions ($c=S,D$) can be obtained as
\cite{hen94,ind96,lak97}
\begin{equation}
\label{eq:couplself}
{\Sigma^{r}_c}_{jk}(E)=\sum_{p,q\in c}
{t_c}_{jp}~{G^{r0}_{c}}_{\!\!\!pq}(E)~{t_c}_{qk},
\end{equation}
with the isolated contact Green's function $G^{r0}_c$
and contact-channel hopping terms $t_c$.

The evaluation of the Coulomb selfenergy $\Sigma^{r}_{ee}$
requires a suitable approximation
scheme due to the infinite Green's function hierarchy (which is a consequence
of the two-particle interaction).
As a first-order expansion (Hartree-Fock diagrams),
four-point Green's functions can be factorized into
linear combinations of products of two-point functions.\cite{hen94,schaefer}
Using this approximation, the Coulomb selfenergy reads
\begin{equation}
\label{eq:coulself}
{\Sigma^{r}_{ee}}_{ml}=\sum_{j,k}\left(V_{mjkl}-V_{jmkl}\right)
\hat{\rho}_{jk}.
\end{equation}
Note that $\Sigma^{<}_{ee}=0$, and
$\Sigma^{r}_{ee}$ is non-local, hermitian and energy-independent (static)
within the considered approximation scheme;
compare also with Ref.~\onlinecite{hen94}.
For convenience, the Hartree potential
(first $V$ term in Eq.~(\ref{eq:coulself}))
\begin{equation}
\label{eq:hartree}
V_H(x)=\frac{1}{\epsilon_0\epsilon_{ch} S}\sum_{\sigma'}\int\!\!dx'
~v(x,x')~\rho_e(x'\!,\sigma')
\end{equation}
can be separated from the retarded Coulomb selfenergy
(compare Eq.~(\ref{eq:vpot})), where the electron charge density $\rho_e$
is given by Eq.~(\ref{eq:erho}).
Hence, the total electrostatic potential of the system reads
$V=V_{dop}+V_H+V_{ext}$.

For integer-number electron filling conditions,
Eq.~(\ref{eq:coulself}) provides
an excellent description of the system for application-relevant temperatures.
However, under nonequilibrium conditions, one has to deal with
non-integer average filling situations, which are beyond the scope of a
first-order (mean-field) selfenergy in general.
In the following section, we will therefore present a multi-configurational
approach which is able to cope with such particle-number fluctuations.

\section{\label{sec:MCSCG}Multi-configurational self-consistent
Green's function}

A thermodynamic state of the transistor channel with fluctuating electron number
can be considered as a weighted mixture of many-body states with integer filling
(configurations) of relevant single-particle states.
For a given $G^<$,
relevant single-particle states are defined as eigenstates of
$\hat{\rho}$ (Eq.~(\ref{eq:densma}))
that exhibit significant occupation fluctuations and
have a sufficiently small dephasing (due to the contact-coupling).
This projection to a relevant single-particle subspace
of dimension $N$ reduces
the resulting Fock subspace dimension $2^N$ significantly,
rendering this approach numerically feasible.

For each configuration,
the Coulomb selfenergy approximation Eq.~(\ref{eq:coulself})
becomes adequate.
Then the Green's function can be written as a configuration-average:
\begin{equation}
\label{eq:gbar}
\bar{G}=\sum_{\kappa} w_{\kappa}~G[\hat{\rho}_{\kappa}],
\end{equation}
where $\hat{\rho}_{\kappa}$
denotes the single-particle density-matrix (derived from $\hat{\rho}$)
for configuration $\kappa$ with weight $w_{\kappa}$.
$G[\hat{\rho}_{\kappa}]$
is the corresponding Green's function (retarded and lesser)
which is obtained by using Eqs.~(\ref{eq:dyson}),~(\ref{eq:coulself}).

The weight vector $w$ defines a projected
nonequilibrium many-body statistical operator in the relevant Fock basis.
(Note that the configurations defined above might not be exact eigenstates of
the projected many-body Hamiltonian, containing Eq.~(\ref{eq:hamopc})
in particular.
In the following, we restrict ourselves to the dominant
diagonal elements of the many-body Hamiltonian in the relevant Fock basis.)
Consequentely, the resulting many-body lesser Green's function reads
\begin{eqnarray}
\label{eq:glessmb}
{G^<_{MB}[w]}_{jk}(t)&=&i~\sum_{\kappa,\lambda}
w_{\kappa}~e^{\frac{i}{\hbar}(E_{\lambda}-E_{\kappa})t}
\\\nonumber
&&\times~
\langle\kappa|c^{\dagger}_k|\lambda\rangle~
\langle\lambda|c_j|\kappa\rangle,
\end{eqnarray}
where $|\kappa\rangle$ denotes a relevant Fock state with energy $E_{\kappa}$.
In principle, $w$ must be chosen
such that $\Delta(G^<_{MB}[w],G^<)=\text{min}$ for a given $G^<$
within the relevant subspace,
where $\Delta$ measures the cumulative difference of spectral weights
of corresponding resonances.
However, for most applications it is sufficient to consider
a vector $w$ that maximizes the entropy
at an effective temperature $T^*(T,V_G,V_D)$
under the (weaker) subsidiary condition that $\hat{\rho}_{MB}[w]=\hat{\rho}$
within the relevant subspace,
where $\hat{\rho}_{MB}[w]$
denotes the many-body result.\cite{ind00,ind02,ind03}
(Under moderate bias conditions, it is justified to assume $T^*\approx T$.)

In turn, $\bar{G}^<$ from Eq.~(\ref{eq:gbar}) can be taken as a new $G^<$,
serving as an input for the calculation scheme described above.
This defines a self-consistency procedure for $\bar{G}$ and $w$,
which we refer to as the
multi-configurational self-consistent Green's function algorithm (MCSCG).
Such an approach resembles the multi-configurational self-consistent field
(MCSCF) approximation.\cite{MCSCF}
However, MCSCG deals with grand-canonical nonequilibrium states and considers
an incoherent superposition (mixture) of different configurations.
Obviously, coherent superpositions of many-body states of varying
particle numbers would be subject to strong dephasing due to the Coulomb
interaction and the resulting entanglement with the environment.

Having solved the many-body diagonalization problem of relevant states,
it is straight-forward to employ this approach
to calculate higher-order correlation functions
(within the relevant subspace).
Note that the MCSCG can also be interpreted as a means to
construct a non-static $\bar{\Sigma}_{ee}$ for Eq.~(\ref{eq:dyson}).

\section{\label{sec:example}Example: SB-FET}

In the following illustrative example, we consider a one-band nanowire-FET
with Schottky-barrier injectors,
having one localized orbital (with two spin directions) per site.
Therefore, only Coulomb matrix elements of the form $V_{ijji}$ are remaining.
Furthermore, we assume nearest-neighbor hopping
with a real hopping parameter $t=\hbar^2/(2m^*a^2)$.
We have used the following device parameters:
The nominally undoped channel has a diameter of $d_{ch}=4$nm and
a length of $L=20$nm (implemented as 20 sites with a spacing of $a=1$nm).
The channel with $\epsilon_{ch}=15$ is surrounded by a gate oxide with
$d_{ox}=10$nm and $\epsilon_{ox}=3.9$,
yielding $\lambda\approx 3.7$nm.
We assume an effective electron mass of $m^*=0.05m_e$ (giving $t=0.77$eV).
The metallic source and drain contacts have a
Schottky-barrier height of $\Phi_{SB}=0.5$eV.
For simplicity, the corresponding contact selfenergy
is assumed to be of the form $\Sigma_c\approx -i\Gamma/2$
(within the band at the outer ends of the channel)
with $\Gamma\approx 76$meV.
Note that this parameter has to be chosen to match the
actual metal contact used in an experiment.
However, it is uncritical for the electronic spectrum
and single-electron charging effects.
The system temperature is $T=77$K.
Up to $N=6$ adaptively chosen relevant single-particle states
were taken into account, depending on the applied voltages
(with $V_S\equiv 0$).

\begin{figure}[ht]
\vspace{-0.3cm}
\begin{center}
\includegraphics[width=8cm]{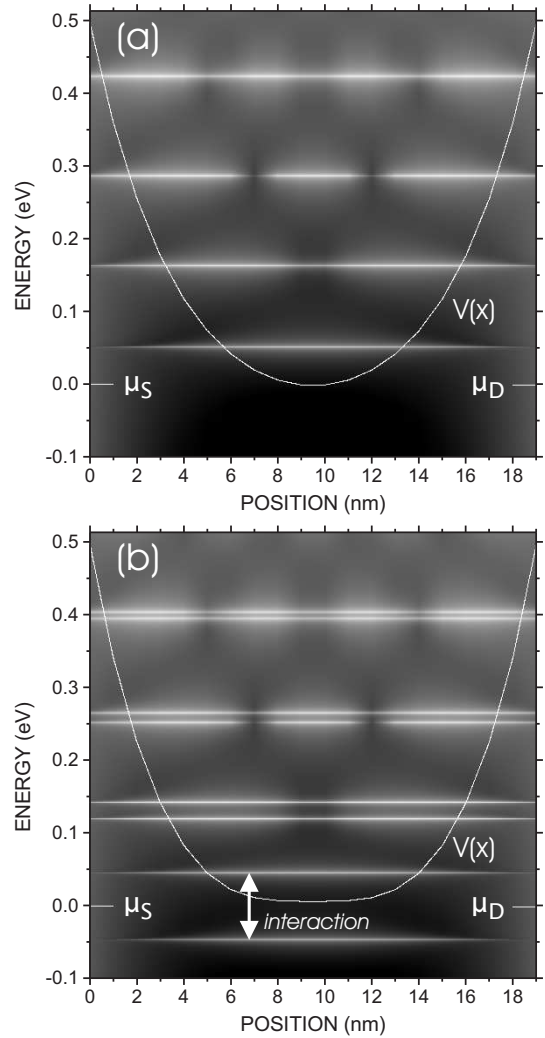}
\end{center}
\vspace{-1.1cm}
\caption{\footnotesize
\label{fig:LDOS}
Spectral function $A(x,E)$ as a grayscale plot for
(a) $V_G=0.59$V and (b) $V_G=0.71$V.
In both cases, $V_D=2$mV is chosen.
The resulting average electron number within the channel is
(a) $N_e\approx 0$ and (b) $N_e\approx 1$
(the electron is located in the lowest resonance).
The solid white line represents the mean-field potential $V(x)$,
whereas $\mu_S$ and $\mu_D$ denote the chemical potentials of
the source and drain contact, respectively.
$T=77$K.
}
\end{figure}

Fig.~\ref{fig:LDOS} visualizes the local density of states (LDOS)
for low drain-source bias conditions
and two different gate voltages $V_G=0.59$V and $V_G=0.71$V,
where the average electron number in the
channel becomes $N_e\approx 0$ and $N_e\approx 1$, respectively.
The existence of quasi-bound states
(i.e. spatially and energetically localized resonances
in the spectral function $A$) yields discrete single-electron energies
with associated interaction energies due to $\Sigma^{r}_{ee}$.
Comparing the situation for $N_e=1$ with $N_e=0$,
the single-electron resonances are moved to higher energies
with respect to the lowest energy state due to the Coulomb repulsion.
Note that each electron is not subject to its own Hartree potential
(see lowest resonance in Fig.~\ref{fig:LDOS}(b)) because
$\Sigma^{r}_{ee}$ does not contain unphysical self-interaction energies,
but includes exchange terms and correctly accounts for the electron spin.
In the shown example,
the next higher available state for a second electron (with opposite spin)
is separated by the interaction energy $V_{00}\approx 93meV$
(see arrow in Fig.~\ref{fig:LDOS}(b)).
In general, energy levels are splitted by exchange energy terms,
which have a significant influence on the energy spectrum.

\begin{figure}[ht]
\vspace{-0.3cm}
\begin{center}
\includegraphics[width=8.5cm]{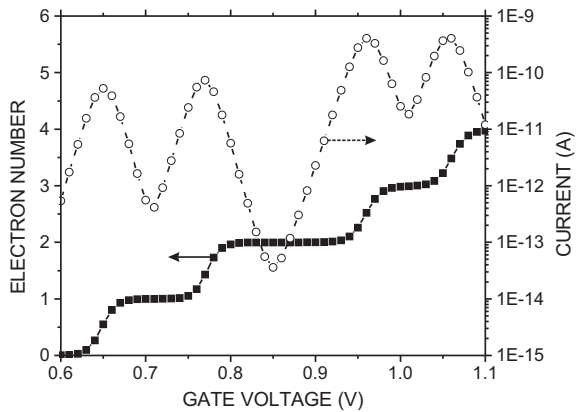}
\end{center}
\vspace{-1.2cm}
\caption{\footnotesize
\label{fig:SETIV}
Single-electron tunneling characteristics for $V_D=2$mV.
The solid line with filled squares
corresponds to the average electron number within
the potential well,
whereas the dashed line with open circles shows
the drain current, exhibiting Coulomb oscillations.
$T=77$K.
}
\end{figure}

As a natural consequence of $h+\Sigma^{r}_{ee}$ we therefore expect to
observe the effect of a step-like electron filling
(under conditions close to equilibrium in particular),
energetically determined by single-electron levels and repulsion energies.
This behavior in fact can be seen in Fig.~\ref{fig:SETIV},
where the electron filling characteristics is plotted for
a varying gate voltage and fixed drain-source bias $V_D=2$mV.
Furthermore, Coulomb oscillations in the accompanying current
through the channel can be identified.

\begin{figure}[ht]
\vspace{-0.3cm}
\begin{center}
\includegraphics[width=8.0cm]{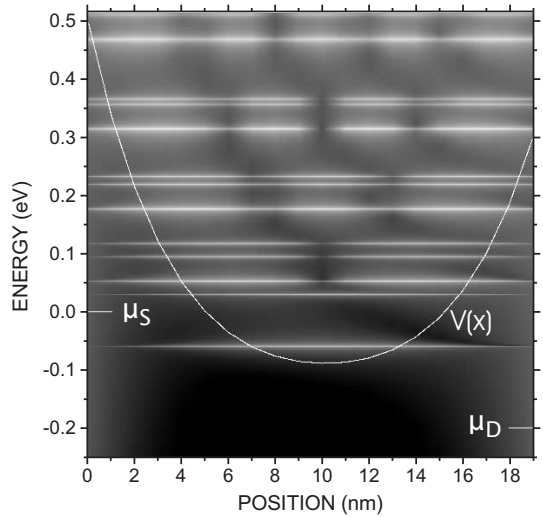}
\end{center}
\vspace{-1.1cm}
\caption{\footnotesize
\label{fig:LDOS2}
Nonequilibrium spectral function $A(x,E)$ as a grayscale plot for
$V_G=0.7$V and $V_D=0.2$V.
The resulting average electron number within the channel is
$N_e\approx 0.22$.
(The solid white line represents the mean-field potential $V(x)$,
whereas $\mu_S$ and $\mu_D$ denote the chemical potentials of
the source and drain contact, respectively.)
$T=77$K.
}
\end{figure}

\begin{figure}[ht]
\vspace{-0.3cm}
\begin{center}
\includegraphics[width=8.5cm]{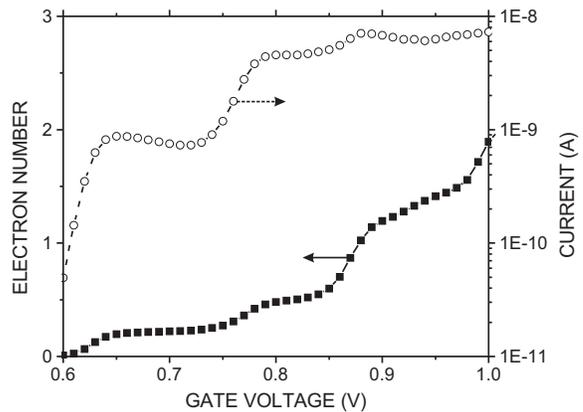}
\end{center}
\vspace{-1.1cm}
\caption{\footnotesize
\label{fig:SETIV2}
Transfer characteristics for $V_D=0.2$V.
The solid line with filled squares
corresponds to the average electron number within
the potential well,
whereas the dashed line with open circles shows the drain current
through the channel.
$T=77$K.
}
\end{figure}

\begin{figure}[ht]
\vspace{-0.3cm}
\begin{center}
\includegraphics[width=8.0cm]{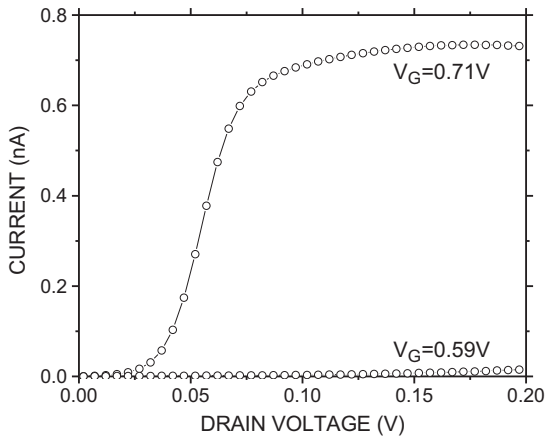}
\end{center}
\vspace{-1.1cm}
\caption{\footnotesize
\label{fig:Output}
Output characteristics for $V_G=0.59$V and $V_G=0.71$V.
$T=77$K.
}
\end{figure}

Models solely based on a selfconsistent Hartree potential
do not provide such quantization effects due to Coulomb repulsion.
With a Hartree approach
(as used with conventional Schr\"odinger-Poisson solvers),
spectral features are solely shifted in energy,
depending on the average electron density.
In contrast, the MCSCG
(as well as the exact diagonalization of the isolated system)
provides a superposition of fading spectra of integer electron numbers
with full interaction energies, however,
having spectral weights that depend on the average filling condition.
The local density of states under nonequilibrium conditions
as shown in Fig.~\ref{fig:LDOS2} clearly demonstrates this behavior,
where the average electron number within the well is $N_e\approx 0.22$.
In fact, the expectation value of the electron number need not be an integer,
especially under non-equilibrium bias conditions, which can be seen
in the corresponding transfer characteristics of the system as
plotted in Fig.~\ref{fig:SETIV2}.
Furthermore, Fig.~\ref{fig:Output} visualizes the output IV characteristics,
where a finite drain voltage is required to pull the chemical potential
of the drain contact below the lowest energy level.
These results clearly demonstrate the strengths of the MCSCG approach,
being able to describe single-electron charging effects under
nonequilibrium bias conditions with fluctuating electron numbers.

In general, we expect the many-body Coulomb interaction
to have a significant impact on the electrical behavior of nano-transistors
if the single-electron charging energy becomes $\geq 4kT$,
having consequences for the transconductance, onset/pinch-off voltages,
sub-threshold currents, and system capacitance.
A more detailed discussion of these aspects will be published elsewhere.

\section{\label{sec:conclu}Conclusion}

The Coulomb Green's function of a one-dimensional FET
in combination with a quantum kinetic description of electronic transport
enables us to describe many-body charging effects within
the transistor channel.
We have presented a multi-configurational self-consistent Green's function
algorithm, which is able to cope with fluctuating electron numbers
under nonequilibrium conditions.
In the discussed example of a nano-FET with Schottky-barrier injectors,
we have visualized how single-electron charging effects arise naturally
as a consequence of the many-body Coulomb repulsion between quasi-bound states.
The usage of a Green's function
formulation permits the systematic extension to further Coulomb diagrams
and the consistent inclusion of phonon scattering.

With the presented theoretical approach, we are able to describe
electronic transport and quantum charging effects in 1D nano-transistors
such as gated carbon nanotubes, semiconductor whiskers,
and 1D CMOS transistors (in coaxial and multi-gate planar geometry).


\end{document}